\documentstyle[aps,prl,preprint]{revtex}
\begin{document}
\draft
\title{Predominantly Superconducting Origin of Large Energy Gaps in \\ 
Underdoped 
Bi$_{2}$Sr$_{2}$CaCu$_{2}$O$_{8-\delta}$ from Tunneling 
Spectroscopy} 
\author{ N.\ Miyakawa$^{1,2}$, J.\ F.\ 
Zasadzinski$^{1,3}$, L.\ Ozyuzer$^{1,4}$, P.\ Guptasarma,$^{1}$ , D.\ 
G.\ Hinks$^{1}$, \\
C.\ Kendziora $^{5}$ and K.\ E.\ Gray $^{1}$ } 
\address{$^{1}$Science and Technology Center for Superconductivity,\\
Materials Science Division, Argonne National Laboratory, Argonne, Illinois 
60439\\
$^{2}$Department of Applied Physics, Science University of Tokyo, 
Kagurazaka 1-3, Shinjuku-ku, Tokyo, 162-8601 Japan\\
$^{3}$Illinois Institute of Technology, Chicago, Illinois 60616\\
$^{4}$Department of Physics, Izmir
Institute of Technology, TR-35210 Izmir, Turkey\\
$^{5}$Naval Research 
Laboratory, Washington, D.C.  20375} \maketitle 
\begin{abstract}
New tunneling data are reported in underdoped 
Bi$_{2}$Sr$_{2}$CaCu$_{2}$O$_{8-\delta}$ using 
superconductor-insulator-superconductor break junctions.  
Energy gaps, $\Delta$, of 51$\pm$2, 54$\pm$2 and 57$\pm$3 meV are 
observed for three 
crystals with T$_{c}$=77, 74, and 70 K respectively.  These energy 
gaps are nearly three times larger than for overdoped crystals with 
similar T$_{c}$. Detailed examination of tunneling spectra over a wide 
doping range from underdoped to overdoped, including the Josephson 
$I_{c}R_{n}$ product, indicate that these energy gaps are 
predominantly of superconducting origin.
\end{abstract} 
\pacs{PACS numbers: 
74.50.+r, 
74.72.Hs, %
74.25.Dw, %
74.62.Dh
} 
Efforts to understand the mechanism of pairing in high-T$_{c}$ superconducting 
(HTS) cuprates are currently focused on the unusual doping dependences of 
superconducting and normal state properties.  In particular, 
underdoped HTS compounds have exhibited pseudogap phenomena above 
T$_{c}$ in both spin and charge excitations.\cite{1} Recently, tunneling\cite{2} 
studies on Bi$_{2}$Sr$_{2}$CaCu$_{2}$O$_{8-\delta}$ 
(Bi2212) in the superconducting state have 
shown a remarkable effect whereby the energy gap exhibits a strong, 
monotonic dependence on doping, increasing substantially in the 
underdoped phase even as T$_{c}$ decreases.  It has been pointed out 
\cite{3} that the 
smooth dependence on doping may nevertheless originate from a 
quasiparticle gap that evolves from superconducting character in the 
overdoped phase to another type (e.g.\ charge density wave, spin 
density wave, etc.) in the underdoped phase.  While the 
measured tunneling gap vs.\  doping is consistent with other 
probes\cite{2} including angle resolved photoemission 
(ARPES)\cite{4,5,6} and Raman\cite{7}, it is at odds with some 
measurements that support a superconducting order parameter scaling 
with T$_{c}$.\cite{8} Thus a critical question is whether 
this relatively large energy gap originates entirely from superconducting 
pairing or has a contribution from some other electronic 
effect.  Here we address 
the nature of the gap measured by tunneling and report new data in 
very underdoped Bi2212 
by superconductor-insulator-superconductor (SIS) break junctions.  
Energy gaps, $\Delta$, of 51$\pm$2, 54$\pm$2 and 57$\pm$3 meV are 
observed for three underdoped crystals with T$_{c}$=77, 74, and 70 K 
respectively, extending the previously reported trend\cite{2} further 
into the underdoped regime.  Detailed examination of the tunneling 
spectra over a wide doping range, including the Josephson $I_{c}R_{n}$ 
product, show that these energy gaps are predominantly of 
superconducting origin.

     Historically, tunneling studies have been relied upon to 
examine the magnitude of the superconducting energy gap 
in both conventional and HTS.\cite{9} In particular, SIS junctions 
provide an accurate measure of 2$\Delta$ from the peak in tunneling 
conductance which is only weakly affected by thermal smearing or 
quasiparticle scattering.\cite{2} However, the large magnitudes of 
energy gaps observed here lead to such extraordinarily large values of $2 
\Delta $/kT$_{c}$ (as high as 20) that it is necessary to examine 
carefully the entire tunneling spectrum to clarify the physical origin 
of these energy gaps.  Most theoretical models of HTS stress the 
importance of electronic correlations such as spin density 
waves\cite{10} and its precursors\cite{11}, or charge density waves 
in the underdoped phase\cite{3,12} which might give rise to 
momentum-dependent quasiparticle excitation gaps, $\Delta _{c}$({\bf 
k}) in addition to those arising from superconductivity, $\Delta 
_{s}$({\bf k}).  In these "two-gap" scenarios the energy gaps often 
add in quadrature\cite{3,12} such that the total energy gap $\Delta$= 
$(\Delta _{s}^{2} + \Delta _{c}^{2})^{1/2}$.  Since these other 
correlation gaps are often used to explain the pseudogap above 
T$_{c}$, our investigation here has a bearing on this issue as well.  
We argue first that if two distinct gaps exist, (1) they should have 
very different doping and temperature dependences; and (2) it is 
unlikely that $\Delta _{s}$({\bf k}) and $\Delta _{c}$({\bf k}) will 
have identical momentum dependences.  Thus the quasiparticle density 
of states (DOS) should exhibit distinct features corresponding to each 
gap.  We observe no evidence of a second gap feature and it will be 
shown that the shape of the gap region spectrum smoothly evolves with 
doping, with features changing mainly in energy scale.

          We also examine a property of SIS junctions that depends 
solely on superconductivity, the Josephson current.  A 
statistical summary of the Josephson $I_{c}R_{n}$ products of over 
40 SIS junctions is presented and it is shown that the largest 
values (both average and maximum) are all found in underdoped 
crystals, where the measured quasiparticle gap is the largest.  
This links the measured quasiparticle gap to a purely 
superconducting energy scale, the Josephson strength.   Thus we 
are forced to conclude that over the range of doping studied 
(from 70 K underdoped to 62 K overdoped) the measured quasiparticle 
gap appears to be due predominantly to superconductivity.

     We grew high quality single crystals using a slightly 
modified floating-zone process as described elsewhere.\cite{2} This yields an 
optimal T$_{c}$ onset of 95 K and the doping is varied through the 
oxygen concentration.  The 70 K underdoped crystal was prepared by a 
different procedure (see ref.\ \cite{13}) and there is good agreement 
among the differently processed samples.  Both SIS break junctions and 
SIN (superconductor-insulator-normal metal) junctions were prepared on 
freshly cleaved surfaces by a point contact technique with Au 
tip.\cite{2,14,15} Tunneling spectra and gap values in the SIN 
junctions\cite{16} are consistent with those presented here but in 
this paper we focus on the SIS junctions.

     In Fig.\ 1 is shown the  dI/dV vs.\ V for an SIS junction on 
the most underdoped crystal with T$_{c}$ = 70 K.  The conductance data have 
been normalized by a constant which is the conductance at V= 340 mV.  
The shape of the conductance is similar to that found on optimally 
doped crystals\cite{2} exhibiting sharp conductance peaks (eV$_{p}$ = 
2$\Delta$), subgap conductance and pronounced dip features at eV= 
3$\Delta$.  At zero bias there is a small Josephson current in the I(V) 
curve (inset of Fig.\  1) which shows up as a conductance peak.  The 
dashed line of Fig.\  1 is a fit using a weighted, momentum averaged 
$d$-wave density of states (DOS).\cite{17} The fit is good except for the 
obvious discrepancies at the dips, and the weighting factor indicates 
that there is preferential tunneling along the ($\pi$,0) point, the 
maximum of the $d$-wave gap.  This analysis gives $\Delta$ = 60 meV 
for the maximum $d$-wave gap and $\Gamma$ = 6 meV where $\Gamma$ is a 
quasiparticle scattering rate.  The data of Fig.\ 1 can also be 
adequately fit with a smeared BCS DOS leading to $\Delta$=57 meV, 
which is exactly half of conductance peak voltage, V$_{p}$.  Thus far, 
reproducibility on this crystal is limited to four separate SIS 
junctions, but in each case a well-defined energy gap is found with 
$\Delta$= 57$\pm$3 meV.  The large energy gaps found here extends the 
previously reported trend [2] further into the underdoped regime and 
leads to a value of 2$\Delta$/kT$_{c}$= 20.

          In Fig.\ 2 we show representative SIS tunneling spectra 
for a wide doping range from overdoped with a T$_{c}$ = 62 K to 
underdoped with a T$_{c}$ = 70 K.  The Josephson peak has been removed 
for clarity.  Four of the curves in Fig.\  2 have been published 
previously\cite{2,14} but are included to display the trend with doping.  To 
compare the spectra, which have widely different gap values, we 
rescale the voltage axis by V$_{p}$/2.  In this way the voltage scale 
is approximately in units of $\Delta$/e.  The bottom three curves of 
Fig.\  2 are new results of 
this study and in the cases of the other underdoped crystals (T$_{c}$ = 74 
K, 77 K) the curves are representative of many different junctions 
formed on each crystal.  Well-defined gap structure was reproducibly 
observed with energy gaps, $\Delta$, of 51$\pm$2, and 54$\pm$2 for the 
crystals with T$_{c}$=77 and 74 K respectively.  What is observed in 
Fig.\ 2 is a smooth evolution of the spectra from overdoped to 
underdoped, with a single gap feature that grows monotonically as the 
doping level is reduced.  All of the curves exhibit subgap 
conductance, which as shown in Fig.\ 1, can be attributed to a 
$d$-wave DOS.  A notable feature in the 
SIS junctions of Fig.\ 2 is the pronounced dip structure\cite{14} which 
remains at eV$\sim$3$\Delta$ over the entire range of this study, i.e.  
$\Delta$= 15 meV - 60 meV.  In none of the curves is there seen any 
evidence of a second peak in the conductance which would be a clear 
indication of another gap in the quasiparticle spectrum.  Rather, what 
is most striking is that the entire gap region spectrum has nearly the 
same shape over the entire doping range, and all that is changing is 
the energy scale of the spectral features.  Thus we find no evidence 
that the nature of the energy gap is changing over the doping range.

     The $I_{c}R_{n}$ value for the junction in Fig.\ 1 is about 2 mV 
as obtained from the I(V) curve, but values as large as 14 mV 
have been observed for one of the other junctions of this same 
crystal.  Here $R_{n}$ is estimated from the high bias conductance 
which is relatively constant as shown in Fig.\ 2.  Large values of 
$I_{c}R_{n}$ (15 mV-25 mV) were previously reported for an 83 K 
underdoped sample.\cite{2} Table I shows the average and maximum 
$I_{c}R_{n}$ values for over 40 SIS junctions on Bi2212 for a variety 
of doping levels.  We find that the average $I_{c}R_{n}$ increases 
with decreased doping and that the three highest values among all the 
junctions are found in underdoped samples, consistent with the large 
quasiparticle gaps observed.  Although the statistical distribution is 
still rough at present, the trend indicates that the quasiparticle gap 
is linked to the Josephson strength, $I_{c}R_{n}$, a purely superconducting 
energy scale.

     The temperature dependence of tunneling conductance was 
measured in some cases and in Fig.\ 3 are shown the results for 
another junction on the T$_{c}$ =77 K underdoped crystal.  The high bias 
junction resistance is $\sim$ 11 k$\Omega$ and all of the data have been 
normalized by this constant value.  Here the Josephson peak at zero 
bias is left in.  As clearly seen in this figure, the superconducting 
gap peak, V$_{p}$, changes very little up to 50 K, but for T $>$ 60 K the 
magnitude of the superconducting gap starts decreasing and states at 
Fermi level start filling in.  The quasiparticle peak coming from 
superconducting gap and the zero-bias peak coming from Josephson 
current continuously disappear near the bulk T$_{c}$.  This is 
important because it means that the local T$_{c}$ of the junction is 
essentially the same as the bulk T$_{c}$.  The decrease of the gap 
magnitude with temperature is also seen directly in the raw data for 
an 83 K underdoped and a 95 K optimal doped crystal\cite{2} and is consistent 
with other SIS break junctions on Bi2212.\cite{18}

     To attempt a more quantitative analysis, the superconducting 
gap, $\Delta$(T), and quasiparticle scattering rate, $\Gamma$(T), as a function 
of temperature have been estimated by fitting the data in Fig.\  3 to a 
simple model for SIS junctions\cite{2,9} which uses a smeared BCS DOS to 
describe the Bi2212.  As discussed earlier in the examination of Fig.\  
1 this analysis might lead to a minor discrepancy in the magnitude of 
the gap when compared to a $d$-wave model, but the simpler smeared BCS 
function makes the analysis much easier.  The results of this 
procedure are shown in Fig.\ 4.  The principal result is that the gap 
magnitude decreases significantly as T increases near T$_{c}$.  For 
T$>$72 K the conductance data are so smeared out that no accurate values 
for $\Delta$ and $\Gamma$ can be obtained.  The decrease in gap 
magnitude near T$_{c}$ is in disagreement with the interpretation of 
recent STM experiments on underdoped Bi2212 where the raw data seem to 
suggest a T-independent gap.\cite{19} We note, however, that $\Delta$(T) 
cannot be inferred directly in SIN junctions due to importance of 
fermi factors in the tunneling conductance.  We estimate the maximum 
Josephson current $I_{c}^{*}$ from the measured conductance peak at 
zero bias and use the junction resistance $R_{n}$ at high bias to plot 
the temperature dependence of the Josephson strength, 
$I_{c}^{*}R_{n}$, normalized to the value obtained at T = 4.2 K.  The 
normalized Josephson strength plotted in Fig.\ 4 is 
approximate and it is used only to estimate the T$_{c}$ of the 
junction which can be seen is very close to the measured bulk T$_{c}$ 
of the crystal (77 K).  We thus can say that the large energy gaps in 
our underdoped crystals correspond to the measured bulk T$_{c}$ and 
are not a consequence of some local deviation in stoichiometry.  

Above the junction T$_{c}$ the quasiparticle gap feature has essentially 
disappeared and only a very weak depression in the conductance at zero 
bias remains.  This behavior is consistent with our previous 
T-dependent SIS data\cite{2} on an 83 K underdoped crystal and a 95 K 
optimally doped crystal as well as other SIS data in the 
literature.\cite{18} We thus find no clear evidence of a pseudogap 
above T$_{c}$ in the SIS tunneling data, again in contrast to the 
strong gap feature observed in recent STM experiments.\cite{19} One 
possible explanation is that the T-dependent pseudogap is highly 
anisotropic in k-space as is indicated in ARPES.\cite{1,6} Since the 
SIS tunneling probes a momentum averaged DOS as indicated by the 
strong subgap conductance in Figs.\ 1 and 2, then the effect of a 
highly anisotropic pseudogap on these junctions may give just a weak 
depression in conductance as is found.
      
     We now summarize the doping dependence of the tunneling data. The shape 
of the SIS quasiparticle spectra are qualitatively the same for all doping 
levels, exhibiting a single gap feature at eV=2$\Delta$ that evolves smoothly 
with doping. The full gap-region spectrum scales with $\Delta$ 
indicating that the 
character of the energy gap is the same over the doping range studied.     
Josephson currents are found in all the SIS junctions and in some cases the 
magnitudes of the $I_{c}R_{n}$ product are very large, nearly 25\% of $\Delta$/e.  
Furthermore, the three highest $I_{c}R_{n}$ values among more than 40 SIS 
junctions were all found in underdoped samples where the quasiparticle 
gap is the largest.  These results indicate that the quasiparticle gaps 
are predominantly due to superconductivity and we find no evidence of 
another gap, $\Delta _{c}$({\bf k}), at low temperatures.  The 
trend of $\Delta$ vs.\  doping closely follows the doping dependence of T$^{*}$, 
the pseudogap temperature for Bi2212.\cite{20} Thus the large gaps in the 
underdoped region suggests that significant superconducting pairing 
correlations exist at temperatures between T$_{c}$ and T* and that 
T$_{c}$ is the temperature where long-range phase coherence sets 
in.\cite{1,21,22}

     The temperature dependence of the superconducting gap, 
$\Delta$(T), in underdoped crystals may also be providing a subtle clue 
about the nature of the superconducting fluctuations above T$_{c}$.  While 
$\Delta$ does not close at T$_{c}$, there is nevertheless a significant 
decrease in its magnitude that is seen even in the raw data.  This 
argues against a picture of tightly bound, pre-formed bosons below T*, 
which would be expected to have a T-independent gap near T$_{c}$.  Rather 
the data are more consistent with intermediate coupling regime 
scenarios for the pairing fluctuations which persist up to T*.\cite{1,22}

This work was partially supported by U.S. Department of Energy, 
Division of Basic Energy Science-Material Science under contract 
No. W-31-109-ENG-38, and the National Science Foundation, 
Office of Science and Technology Centers under contract No. 
DMR 91-20000.

\newpage

 Fig.\  1 Differential conductance, dI/dV, for SIS break junction on 
an underdoped single crystal of Bi2212 with T$_{c}$=70 K.  The inset 
shows a Josephson current at zero bias.  

Fig.\  2 Normalized SIS 
tunneling conductances of Bi2212 with various doping levels from 
underdoped to overdoped.  The voltage axis has been rescaled in units 
of $\Delta$/e.  

Fig.\  3 Temperature dependence of SIS tunneling 
conductance on an underdoped Bi2212 (T$_{c}$=77 K) break junction.  
For clarity, each conductance has been normalized by its value at 200 
mV and (except for the 5 K curve) is offset vertically.  

Fig.\  4 Temperature dependence of superconducting gap $\Delta$(T) (circle), 
quasiparticle scattering rate $\Gamma$(T) (triangle) and normalized 
Josephson strength, $I_{c}^{*}R_{n}$ (square) (see text in details), 
where the normalization of $I_{c}^{*}R_{n}$(T) has been done by 
$I_{c}^{*}R_{n}$(4.2 K).  The full curve represents the BCS curve of 
superconducting gap $\Delta$(T).
\end{document}